\documentclass[twocolumn,english,conference]{IEEEtran}
\usepackage[T1]{fontenc}
\usepackage[latin9]{inputenc}
\usepackage{babel}
\usepackage{amsmath}
\usepackage{amssymb}
\usepackage{graphicx}
\usepackage[unicode=true,
 bookmarks=true,bookmarksnumbered=true,bookmarksopen=true,bookmarksopenlevel=1,
 breaklinks=false,pdfborder={0 0 0},backref=false,colorlinks=false]
 {hyperref}
\hypersetup{pdftitle={Your Title},
 pdfauthor={Your Name},
 pdfpagelayout=OneColumn,pdfnewwindow=true,pdfstartview=XYZ,plainpages=false}
\usepackage{breakurl}

\makeatletter
\usepackage{babel}

\makeatother

\begin{document}

\title{Sequential Inverse Approximation of a Regularized Sample Covariance
Matrix}

\author{Tomer Lancewicki\\
EBay Inc.\\
625 6th Ave\\
New York, NY\\
Email: tlancewicki@ebay.com}
\maketitle
\begin{abstract}
One of the goals in scaling sequential machine learning methods pertains
to dealing with high-dimensional data spaces. A key related challenge
is that many methods heavily depend on obtaining the inverse covariance
matrix of the data. It is well known that covariance matrix estimation
is problematic when the number of observations is relatively small
compared to the number of variables. A common way to tackle this problem
is through the use of a shrinkage estimator that offers a compromise
between the sample covariance matrix and a well-conditioned matrix,
with the aim of minimizing the mean-squared error. We derived sequential
update rules to approximate the inverse shrinkage estimator of the
covariance matrix. The approach paves the way for improved large-scale
machine learning methods that involve sequential updates. 
\end{abstract}

\section{Introduction}

The covariance matrix of multivariate data is required in many sequential
machine learning and neural-networks (NN) based applications \cite{icml2014_1},
including speech recognition \cite{DML3}, deep learning architectures
for image processing and computer vision \cite{Arel2,Deep3,Lancewicki2014382},
stochastic fuzzy NN's \cite{DML4}, pricing option contracts in financial
markets \cite{DML10}, adaptive tracking control problems \cite{DML9},
detection tasks \cite{DML2}, reinforcement learning \cite{lancewicki_RL},
and many others.

In settings where data arrives sequentially, the covariance matrix
is required to be updated in an online manner \cite{lancewicki2015sequential,DML8}.
Techniques such as cross-validation, which attempt to impose regularization,
or model selection are typically not feasible in such settings \cite{DML11}.
Instead, to minimize complexity, it is often assumed that the covariance
matrix is known in advance \cite{DML4} or that it is restricted to
a specific simplified structure, such as a diagonal matrix \cite{DML6,Arel2}.
Moreover, when the number of observations $n$ is comparable to the
number of variables $p$ the covariance estimation problem becomes
far more challenging. In such scenarios, the sample covariance matrix
is not well-conditioned nor is it necessarily invertible (despite
the fact that those two properties are required for most applications).
When $n\leq p$, the inversion cannot be computed at all \cite{MatrixHandbook,lancewicki2017regularization}.

An extensive body of literature concerning improved estimators in
such situations exists \cite{bickel,rohde2011}. However, in the absence
of a specific knowledge about the structure of the true covariance
matrix, the most successful approach so far has, arguably, been shrinkage
estimation \cite{ledoit2011nonlinear}. It has been demonstrated in
\cite{slda8} that the largest sample eigenvalues are systematically
biased upward, and the smallest ones downward. This bias is corrected
by pulling down the largest eigenvalues and pushing up the smallest
ones, toward their grand mean.

The optimal solution of the shrinkage estimator is solved analytically,
which is a huge advantage for deep learning architectures, since a
key factor in realizing such architectures is the resource complexity
involved in their training \cite{erhan2009difficulty}. An example
of such deep architecture is the deep spatiotemporal inference network
(DeSTIN) \cite{Arel2}. The latter extensively utilizes the \emph{quadratic
discriminant analysis} (QDA) classifier under the simplified assumption
that the covariance matrices involved in the process are diagonal.
Such assumption is made in order to avoid additional complexity during
the training and inference processes. It is well known that for a
small ratio of training observations $n$ to observation dimensionality
$p$, the QDA classifier performs poorly, due to highly variable class
conditional sample covariance matrices. In order to improve the classifiers'
performance, regularization is required, with the aim of providing
an appropriate compromise between the bias and variance of the solution.
It have been demonstrated in \cite{Lance,lancewicki2014dimensionality}
that the QDA classifier can be improved tremendously using shrinkage
estimators. The sequential approximated inverse of the shrinkage estimator,
derived in this paper, allows us to utilize the shrinkage estimator
in the DeSTIN architecture with relatively negligible additional complexity
to the architecture. In addition, the relatively simple update rules
pave the way to implement the inverse shrinkage estimator on analog
computational circuits, offering the potential for large improvement
in power efficiency \cite{Arel1}. 

The rest of this paper is organized as follows: Section 2 presents
the general idea of the shrinkage estimator. In Section 3, we derived
a sequential update for the shrinkage estimator, while in Section
4, the related approximated inverses are derived. In Section 5, we
conduct an experimental study and examine the sequential update rules.

Notations: we denote vectors in lowercase boldface letters and matrices
in uppercase boldface. The transpose operator is denoted as $\left(\cdot\right)^{T}$.
The trace, the determinant and the Frobenius norm of a matrix are
denoted as $\mathrm{Tr}\left(\cdot\right)$, $\left|\cdot\right|$
and $\left\Vert \cdot\right\Vert _{F}$, respectively. The identity
matrix is denoted as $\mathbf{I}$, while $\mathbf{e}=\left[1,1,\ldots,1\right]^{T}$
is a column vector of all ones. For any real matrices $\mathbf{R}_{1}$
and $\mathbf{R}_{2}$, the inner product is defined as $\left\langle \mathbf{R}_{1},\mathbf{R}_{2}\right\rangle =\mathrm{Tr}\left(\mathbf{R}_{1}^{T}\mathbf{R}_{2}\right)$,
where $\left\langle \mathbf{R}_{1},\mathbf{R}_{1}\right\rangle =\left\Vert \mathbf{R}_{1}\right\Vert _{F}^{2}$
\cite[Sec. 2.20]{MatrixHandbook}.

\section{Shrinkage Estimator for Covariance Matrices}

We briefly review a single-target shrinkage estimator by following
\cite{slda8,slda12}, which is generally applied to high-dimensional
estimation problems. Let $\left\{ \mathbf{x}_{i}\right\} _{i=1}^{n}$
be a sample of \emph{independent identically distributed} (i.i.d.)
$p$-dimensional vectors drawn from a density with a mean $\boldsymbol{\mu}$
and covariance matrix $\boldsymbol{\Sigma}$. When the number of observations
$n$ is large (i.e., $n\gg p$), the most common estimator of $\boldsymbol{\Sigma}$
is the sample covariance matrix 
\begin{equation}
\mathbf{S}_{n}=\frac{1}{n-1}\sum_{i=1}^{n}\left(\mathbf{x}_{i}-\mathbf{m}_{n}\right)\left(\mathbf{x}_{i}-\mathbf{m}_{n}\right)^{T},\label{eq:Smatrix}
\end{equation}
where $\mathbf{m}_{n}$ is the sample mean, defined as 
\begin{equation}
\mathbf{m}_{n}=\frac{1}{n}\sum_{i=1}^{n}\mathbf{x}_{i}.\label{eq:mean}
\end{equation}
Both $\mathbf{S}_{n}$ and $\mathbf{m}_{n}$ are unbiased estimators
of $\boldsymbol{\Sigma}$ and $\boldsymbol{\mu}$, respectively, i.e.,
$E\left\{ \mathbf{S}_{n}\right\} =\boldsymbol{\Sigma}$ and $E\left\{ \mathbf{m}_{n}\right\} =\boldsymbol{\mu}$.
The shrinkage estimator $\hat{\mathbf{\boldsymbol{\Sigma}}}\left(\lambda_{n}\right)$
is in the form 
\begin{equation}
\hat{\mathbf{\boldsymbol{\Sigma}}}\left(\lambda_{n}\right)=(1-\lambda_{n})\mathbf{S}_{n}+\lambda_{n}\mathbf{T}_{n}\label{eq:estimator}
\end{equation}
where the target $\mathbf{T}_{n}$ is a restricted estimator of $\boldsymbol{\Sigma}$
defined as 
\begin{equation}
\mathbf{T}_{n}=\frac{\mathrm{Tr}\left(\mathbf{S}_{n}\right)}{p}\mathbf{I}.\label{eq:Tar1}
\end{equation}
The work in \cite{slda8} proposed to find an estimator $\hat{\mathbf{\boldsymbol{\Sigma}}}\left(\lambda_{n}\right)$
which minimizes the \emph{mean squared error} (MSE) with respect to
$\lambda_{n}$, i.e., 
\begin{equation}
\lambda_{On}=\arg\min_{\lambda_{n}}E\left\{ \left\Vert \hat{\mathbf{\boldsymbol{\Sigma}}}\left(\lambda_{n}\right)-\boldsymbol{\Sigma}\right\Vert _{F}^{2}\right\} \label{eq:lambdaO-3}
\end{equation}
and can be given by the distribution-free formula 
\begin{equation}
\lambda_{On}=\frac{E\left\{ \left\langle \mathbf{T}_{n}-\mathbf{S}_{n},\boldsymbol{\Sigma}-\mathbf{S}_{n}\right\rangle \right\} }{E\left\{ \left\Vert \mathbf{T}_{n}-\mathbf{S}_{n}\right\Vert _{F}^{2}\right\} }.\label{eq:lambdaO}
\end{equation}
The scalar $\lambda_{On}$ is called the oracle shrinkage coefficient,
since its depends on the unknown covariance matrix $\boldsymbol{\Sigma}$.
Therefore, $\lambda_{On}$ \eqref{eq:lambdaO} must be estimated.
The latter can be estimated from its sample counterparts as in \cite{slda12}.
We denote this estimator as $\hat{\lambda}_{On}$.

\section{Sequential Update of the Shrinkage Estimator}

We want to know what happens to $\hat{\mathbf{\boldsymbol{\Sigma}}}\left(\lambda_{n}\right)$
\eqref{eq:estimator} when we add an observation $\mathbf{x}_{n+1}$,
using only the current knowledge of $\mathbf{S}_{n}$, $\mathbf{m}_{n}$
and $n$. Setting $\mathbf{d}_{n+1}=\mathbf{x}_{n+1}-\mathbf{m}_{n}$
while using \cite[15.12.(c)]{MatrixHandbook}, we have the following
update rules for $\mathbf{m}_{n}$ \eqref{eq:mean} and $\mathbf{S}_{n}$
\eqref{eq:Smatrix} when an observation $\mathbf{x}_{n+1}$ is added
\begin{equation}
\mathbf{m}_{n+1}=\mathbf{m}_{n}+\frac{1}{n+1}\mathbf{d}_{n+1}
\end{equation}
\begin{equation}
\mathbf{S}_{n+1}=\frac{n-1}{n}\mathbf{S}_{n}+\frac{1}{n+1}\mathbf{d}_{n+1}\mathbf{d}_{n+1}^{T}.\label{eq:Scov}
\end{equation}

Based on $\mathbf{S}_{n+1}$ \eqref{eq:Scov}, we can write the update
rule for the target $\mathbf{T}_{n}$ \eqref{eq:Tar1} as

\begin{equation}
\mathbf{T}_{n+1}=\frac{n-1}{n}\mathbf{T}_{n}+\frac{1}{\left(n+1\right)p}\left\Vert \mathbf{d}_{n+1}\right\Vert _{F}^{2}\mathbf{I}\label{eq:Tar1-1}
\end{equation}
By using $\mathbf{S}_{n+1}$ \eqref{eq:Scov} and $\mathbf{T}_{n+1}$
\eqref{eq:Tar1-1}, the update rule for the shrinkage estimator $\hat{\mathbf{\boldsymbol{\Sigma}}}\left(\lambda_{n}\right)$
\eqref{eq:estimator} can be written as 
\begin{equation}
\hat{\mathbf{\boldsymbol{\Sigma}}}\left(\lambda_{n+1}\right)=\mathbf{G}_{n}+\mathbf{F}_{n}\label{eq:sigupdate}
\end{equation}
where $\mathbf{G}_{n}$ and $\mathbf{F}_{n}$ defined as 
\begin{equation}
\mathbf{G}_{n}=\frac{n-1}{n}\hat{\mathbf{\boldsymbol{\Sigma}}}\left(\lambda_{n}\right)+\left(1-\lambda_{n+1}\right)\frac{1}{n+1}\mathbf{d}_{n+1}\mathbf{d}_{n+1}^{T}\label{eq:G1}
\end{equation}
and 
\[
\mathbf{F}_{n}=\frac{1}{\left(n+1\right)p}\lambda_{n+1}\left\Vert \mathbf{d}_{n+1}\right\Vert _{F}^{2}\mathbf{I}
\]
\begin{equation}
+\frac{n-1}{n}(\lambda_{n}-\lambda_{n+1})\left(\mathbf{S}_{n}-\mathbf{T}_{n}\right),\label{eq:F}
\end{equation}
respectively. Based on the above update rules, we derive the sequential
update rules for the inverse of the shrinkage estimator.

\section{Sequential Update for the Inverse of the Shrinkage Estimator}

In this section, we derived approximated inverses of the shrinkage
estimator which are updated sequentially and do not involve any matrix
inversion. We start, therefore, from the inverse of the sample covariance
matrix $\mathbf{S}_{n+1}$ that can be obtained from the current inverse
of $\mathbf{S}_{n}$ \eqref{eq:Smatrix} using the Sherman-Morrison-Woodbury
matrix identity \cite[Ch. 3]{Duda} as 
\begin{equation}
\mathbf{S}_{n+1}^{-1}=\frac{n}{n-1}\left(\mathbf{S}_{n}^{-1}-\frac{\mathbf{S}_{n}^{-1}\mathbf{d}_{n+1}\mathbf{d}_{n+1}^{T}\mathbf{S}_{n}^{-1}}{\frac{n^{2}-1}{n}+\mathbf{d}_{n+1}^{T}\mathbf{S}_{n}^{-1}\mathbf{d}_{n+1}}\right).\label{eq:sinv}
\end{equation}
The last update rule can be used only if $\mathbf{S}_{n}$ is invertible.
It will not be invertible for $n\leq p$. Since the shrinkage estimator
$\hat{\mathbf{\boldsymbol{\Sigma}}}\left(\lambda_{n}\right)$ \eqref{eq:estimator}
is a regularized version of $\mathbf{S}_{n}$ \eqref{eq:Smatrix},
an inverse exists for any $n$. This inverse of $\hat{\mathbf{\boldsymbol{\Sigma}}}\left(\lambda_{n}\right)$
\eqref{eq:estimator} involves two main steps. The first one is to
update the inverse of $\mathbf{G}_{n}$ \eqref{eq:G1} from an inverse
of $\hat{\mathbf{\boldsymbol{\Sigma}}}\left(\lambda_{n}\right)$ \eqref{eq:estimator}.
The second is to update the next step inverse of $\hat{\mathbf{\boldsymbol{\Sigma}}}\left(\lambda_{n}\right)$
from $\mathbf{F}_{n}$ \eqref{eq:F} and the inverse of $\mathbf{G}_{n}$
\eqref{eq:G1} calculated in the first step. Suppose, for example,
that the exact inverse of $\hat{\mathbf{\boldsymbol{\Sigma}}}\left(\lambda_{n}\right)$
\eqref{eq:estimator}, denoted as $\hat{\mathbf{\boldsymbol{\Sigma}}}^{-1}\left(\lambda_{n}\right)$,
is known. In the same manner as in $\mathbf{S}_{n+1}^{-1}$ \eqref{eq:sinv},
the inverse for $\mathbf{G}_{n}$ \eqref{eq:G1} can be calculated
from $\hat{\mathbf{\boldsymbol{\Sigma}}}^{-1}\left(\lambda_{n}\right)$
as 
\begin{equation}
\begin{array}{c}
\mathbf{G}_{n}^{-1}=\\
\frac{n}{n-1}\left(\hat{\mathbf{\boldsymbol{\Sigma}}}^{-1}\left(\lambda_{n}\right)-\frac{\hat{\mathbf{\boldsymbol{\Sigma}}}^{-1}\left(\lambda_{n}\right)\mathbf{d}_{n+1}\mathbf{d}_{n+1}^{T}\hat{\mathbf{\boldsymbol{\Sigma}}}^{-1}\left(\lambda_{n}\right)}{\frac{n^{2}-1}{n\left(1-\lambda_{n+1}\right)}+\mathbf{d}_{n+1}^{T}\hat{\mathbf{\boldsymbol{\Sigma}}}^{-1}\left(\lambda_{n}\right)\mathbf{d}_{n+1}}\right).
\end{array}\label{eq:Gn}
\end{equation}

Using \cite[15.11.(b)]{MatrixHandbook}, the exact inverse of $\hat{\mathbf{\boldsymbol{\Sigma}}}\left(\lambda_{n+1}\right)$
can be calculated from $\mathbf{G}_{n}^{-1}$ \eqref{eq:Gn} and $\mathbf{F}_{n}$
\eqref{eq:F} with $p$ iterations 
\[
\left(\mathbf{G}_{n}^{\left(i+1\right)}\right)^{-1}=\left(\mathbf{G}_{n}^{\left(i\right)}+\mathbf{f}_{i}\mathbf{e}_{i}^{T}\right)^{-1}
\]
\begin{equation}
=\left(\mathbf{G}_{n}^{\left(i\right)}\right)^{-1}-\frac{\left(\mathbf{G}_{n}^{\left(i\right)}\right)^{-1}\mathbf{f}_{i}\mathbf{e}_{i}^{T}\left(\mathbf{G}_{n}^{\left(i\right)}\right)^{-1}}{1+\mathbf{e}_{i}^{T}\left(\mathbf{G}_{n}^{\left(i\right)}\right)^{-1}\mathbf{f}_{i}},\, i=1,\ldots,p
\end{equation}
where $\mathbf{f}_{i}$ and $\mathbf{e}_{i}$ are the $i$ columns
of $\mathbf{F}_{n}$ \eqref{eq:F} and the identity matrix $\mathbf{I}$,
respectively. The inverse of $\hat{\mathbf{\boldsymbol{\Sigma}}}\left(\lambda_{n+1}\right)$
\eqref{eq:sigupdate} is equal to the output of the last iteration,
i.e., 
\begin{equation}
\hat{\mathbf{\boldsymbol{\Sigma}}}^{-1}\left(\lambda_{n+1}\right)=\left(\mathbf{G}_{n}^{\left(p+1\right)}\right)^{-1}.\label{eq:sigapp}
\end{equation}
In order to avoid the calculation of $p$ iterations, we can use approximations
for $\hat{\mathbf{\boldsymbol{\Sigma}}}^{-1}\left(\lambda_{n+1}\right)$
\eqref{eq:sigapp}. The inverse approximations of the shrinkage estimator
are discussed in the following section.

\subsection{Inverse Approximations for the Shrinkage Estimator}

We consider two approximations for $\hat{\mathbf{\boldsymbol{\Sigma}}}^{-1}\left(\lambda_{n+1}\right)$
\eqref{eq:sigapp}. The first approximation is defined as 
\begin{equation}
\tilde{\boldsymbol{\Sigma}}_{1}^{-1}\left(\lambda_{n+1}\right)=\tilde{\mathbf{G}}_{n}^{-1}-\alpha_{n}\tilde{\mathbf{G}}_{n}^{-1}\mathbf{F}_{n}\tilde{\mathbf{G}}_{n}^{-1}\label{eq:siginvapp}
\end{equation}
where 
\[
\tilde{\mathbf{G}}_{n}^{-1}=
\]
\begin{equation}
\frac{n}{n-1}\left(\tilde{\boldsymbol{\Sigma}}_{1}^{-1}\left(\lambda_{n}\right)-\frac{\tilde{\boldsymbol{\Sigma}}_{1}^{-1}\left(\lambda_{n}\right)\mathbf{d}_{n+1}\mathbf{d}_{n+1}^{T}\tilde{\boldsymbol{\Sigma}}_{1}^{-1}\left(\lambda_{n}\right)}{\frac{n^{2}-1}{n\left(1-\lambda_{n+1}\right)}+\mathbf{d}_{n+1}^{T}\tilde{\boldsymbol{\Sigma}}_{1}^{-1}\left(\lambda_{n}\right)\mathbf{d}_{n+1}}\right).\label{eq:invg}
\end{equation}
The matrix $\tilde{\mathbf{G}}_{n}^{-1}$ \eqref{eq:invg} differs
from $\mathbf{G}_{n}^{-1}$ \eqref{eq:Gn} in the fact that it relies
on the approximated inverse $\tilde{\boldsymbol{\Sigma}}^{-1}\left(\lambda_{n}\right)$
\eqref{eq:siginvapp}, instead of the exact inverse $\hat{\mathbf{\boldsymbol{\Sigma}}}^{-1}\left(\lambda_{n}\right)$
\eqref{eq:sigapp}. A possible motivation to justify the update rule
\eqref{eq:siginvapp} stems from the mean value theorem as explained
in \cite{Miller1981}. Another motivation arises from the Neumann
series \cite[Sec.19.15]{MatrixHandbook} where $\hat{\mathbf{\boldsymbol{\Sigma}}}^{-1}\left(\lambda_{n+1}\right)$
\eqref{eq:sigapp} is approximately equal to $\tilde{\boldsymbol{\Sigma}}^{-1}\left(\lambda_{n+1}\right)$
\eqref{eq:siginvapp} for $\alpha=1$ and relatively small $\mathbf{F}_{n}$.
We define $\alpha_{n}$ as the value that minimizes the reconstruction
squared error, i.e., 
\begin{equation}
\alpha_{n}=\arg\min_{\alpha}\left\Vert \left(\tilde{\mathbf{G}}_{n}^{-1}-\alpha\tilde{\mathbf{G}}_{n}^{-1}\mathbf{F}_{n}\tilde{\mathbf{G}}_{n}^{-1}\right)\hat{\mathbf{\boldsymbol{\Sigma}}}\left(\lambda_{n+1}\right)-\mathbf{I}\right\Vert _{F}^{2}
\end{equation}
and is equal to 
\begin{equation}
\alpha_{n}=\frac{\mathrm{Tr}\left(\tilde{\mathbf{G}}_{n}^{-1}\mathbf{F}_{n}\tilde{\mathbf{G}}_{n}^{-1}\hat{\mathbf{\boldsymbol{\Sigma}}}\left(\lambda_{n+1}\right)\left(\tilde{\mathbf{G}}_{n}^{-1}\hat{\mathbf{\boldsymbol{\Sigma}}}\left(\lambda_{n+1}\right)-\mathbf{I}\right)\right)}{\left\Vert \tilde{\mathbf{G}}_{n}^{-1}\mathbf{F}_{n}\tilde{\mathbf{G}}_{n}^{-1}\hat{\mathbf{\boldsymbol{\Sigma}}}\left(\lambda_{n+1}\right)\right\Vert _{F}^{2}}
\end{equation}
Additional simplification can be taken by looking at the last term
in $\mathbf{F}_{n}$ \eqref{eq:F}. Under the assumption that the
difference $\lambda_{n}-\lambda_{n+1}$ is relatively small, we can
write an approximation for $\mathbf{F}_{n}$ \eqref{eq:F} by neglecting
its last term, i.e., 
\begin{equation}
\tilde{\mathbf{F}}_{n}=\frac{1}{\left(n+1\right)p}\lambda_{n+1}\left\Vert \mathbf{d}_{n+1}\right\Vert _{F}^{2}\mathbf{I}\label{eq:F-1}
\end{equation}
This will lead to the second approximation for $\hat{\mathbf{\boldsymbol{\Sigma}}}^{-1}\left(\lambda_{n+1}\right)$
\eqref{eq:sigapp}, denoted as 
\begin{equation}
\tilde{\boldsymbol{\Sigma}}_{2}^{-1}\left(\lambda_{n+1}\right)=\tilde{\mathbf{G}'}_{n}^{-1}-\alpha'_{n}\tilde{\mathbf{G}'}_{n}^{-1}\tilde{\mathbf{F}}_{n}\tilde{\mathbf{G}'}_{n}^{-1}\label{eq:siginvapp-1}
\end{equation}
where 
\begin{equation}
\begin{array}{c}
\tilde{\mathbf{G}'}_{n}^{-1}=\\
\frac{n}{n-1}\left(\tilde{\boldsymbol{\Sigma}}_{2}^{-1}\left(\lambda_{n}\right)-\frac{\tilde{\boldsymbol{\Sigma}}_{2}^{-1}\left(\lambda_{n}\right)\mathbf{d}_{n+1}\mathbf{d}_{n+1}^{T}\tilde{\boldsymbol{\Sigma}}_{2}^{-1}\left(\lambda_{n}\right)}{\frac{n^{2}-1}{n\left(1-\lambda_{n+1}\right)}+\mathbf{d}_{n+1}^{T}\tilde{\boldsymbol{\Sigma}}_{2}^{-1}\left(\lambda_{n}\right)\mathbf{d}_{n+1}}\right)
\end{array}\label{eq:invg-1}
\end{equation}
and $\alpha'_{n}$ is calculated by 
\begin{equation}
\alpha'_{n}=\frac{\left(n+1\right)p\mathrm{Tr}\left(\tilde{\mathbf{G}'}_{n}^{-2}\hat{\mathbf{\boldsymbol{\Sigma}}}\left(\lambda_{n+1}\right)\left(\tilde{\mathbf{G}'}_{n}^{-1}\hat{\mathbf{\boldsymbol{\Sigma}}}\left(\lambda_{n+1}\right)-\mathbf{I}\right)\right)}{\lambda_{n+1}\left\Vert \mathbf{d}_{n+1}\right\Vert _{F}^{2}\left\Vert \tilde{\mathbf{G}'}_{n}^{-2}\hat{\mathbf{\boldsymbol{\Sigma}}}\left(\lambda_{n+1}\right)\right\Vert _{F}^{2}}
\end{equation}
We examine these two approximations in the following section.

\section{Experiments}

In this section we implement and evaluate the sequential update of
the inverse shrinkage estimator. As in \cite{MMSE}, we assume that
the observations are i.i.d Gaussian vectors. In order to study the
estimators performance, an autoregressive covariance matrix $\boldsymbol{\Sigma}$
is used. We let $\boldsymbol{\Sigma}$ be the covariance matrix of
a Gaussian AR(1) process \cite{MMSE32}, denoted by 
\begin{equation}
\boldsymbol{\Sigma}_{AR}=\left\{ \sigma_{ij}=r^{\left|i-j\right|}\right\} .\label{eq:AR}
\end{equation}
As in \cite{bickel,MMSE}, we use $r=0.5$. In all simulations, we
set $p=50$ and let $n$ range from 1 to 30. Each simulation is repeated
200 times and the average values are plotted as a function of $n$.
The experimental results are summarized in box plots. On each box,
the central mark is the median, the edges of the box are the 25th
and 75th percentiles, and the whiskers correspond to approximately
+/\textendash $2.7\sigma$ or 99.3 coverage if the data are normally
distributed. The outliers are plotted individually.

The reconstruction errors of the approximated inverses $\tilde{\boldsymbol{\Sigma}}_{1}^{-1}\left(\lambda_{n}\right)$
\eqref{eq:siginvapp} and $\tilde{\boldsymbol{\Sigma}}_{2}^{-1}\left(\lambda_{n}\right)$
\eqref{eq:siginvapp-1} are defined by 
\begin{equation}
e_{1}\left(n\right)=\frac{1}{p}\left\Vert \tilde{\boldsymbol{\Sigma}}_{1}^{-1}\left(\lambda_{n}\right)\hat{\mathbf{\boldsymbol{\Sigma}}}\left(\lambda_{n}\right)-\mathbf{I}\right\Vert _{F}^{2},\label{eq:e1n}
\end{equation}
and 
\begin{equation}
e_{2}\left(n\right)=\frac{1}{p}\left\Vert \tilde{\boldsymbol{\Sigma}}_{2}^{-1}\left(\lambda_{n}\right)\hat{\mathbf{\boldsymbol{\Sigma}}}\left(\lambda_{n}\right)-\mathbf{I}\right\Vert _{F}^{2},\label{eq:e2n}
\end{equation}
respectively. These reconstruction errors are normalized with $p$
since it is the squared Frobenius norm of the identity matrix $\mathbf{I}$.
We examine the approximated inverse $\tilde{\boldsymbol{\Sigma}}_{1}^{-1}\left(\lambda_{n}\right)$
\eqref{eq:siginvapp} and $\tilde{\boldsymbol{\Sigma}}_{2}^{-1}\left(\lambda_{n}\right)$
\eqref{eq:siginvapp-1} where $\lambda_{n}$ is equal to $\hat{\lambda}_{On}$
\cite{slda12}. The experimental results for the reconstruction errors
$e_{1}\left(n\right)$ \eqref{eq:e1n} and $e_{2}\left(n\right)$
\eqref{eq:e2n} are summarized in Fig. 1 and Fig. 2, respectively.
The values of $e_{1}\left(n\right)$ \eqref{eq:e1n} converge on average
to zero as the number of observations $n$ increase. In several simulations,
however, the update rule accumulates error and diverges.

\begin{figure}[h]
\vskip 0.2in

\begin{centering}
\centerline{\includegraphics[scale=0.04]{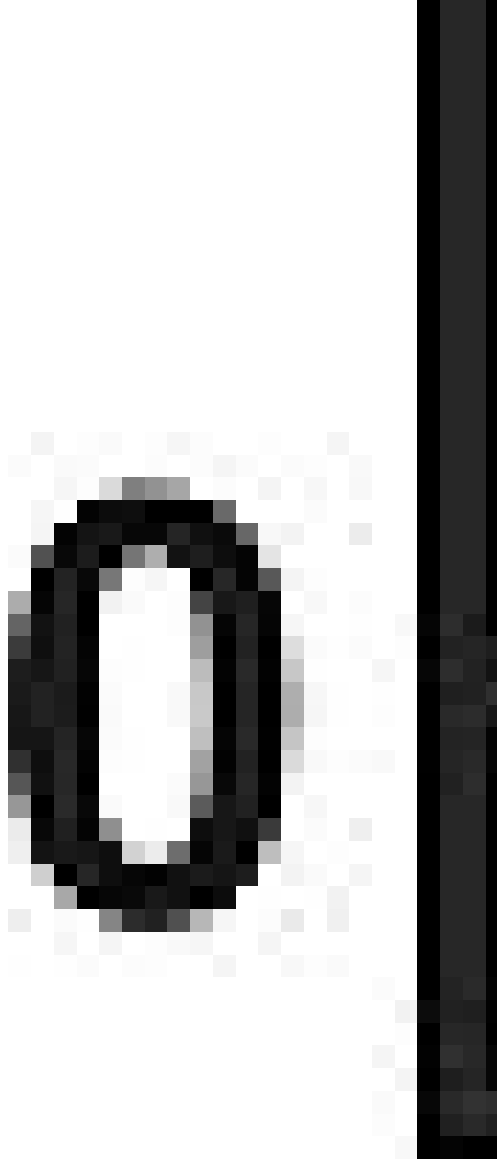}} 
\par\end{centering}

\begin{centering}
\protect\protect\protect\protect\caption{$e_{1}\left(n\right)$ \eqref{eq:e1n} for $\boldsymbol{\Sigma}_{AR}$
\eqref{eq:AR} of AR(1) process with $p=50$ and $r=0.5$. }

\par\end{centering}

\begin{centering}
\label{icml-historical-2-1} 
\par\end{centering}

\vskip -0.2in 
\end{figure}

\begin{figure}[h]
\vskip 0.2in

\begin{centering}
\centerline{\includegraphics[scale=0.04]{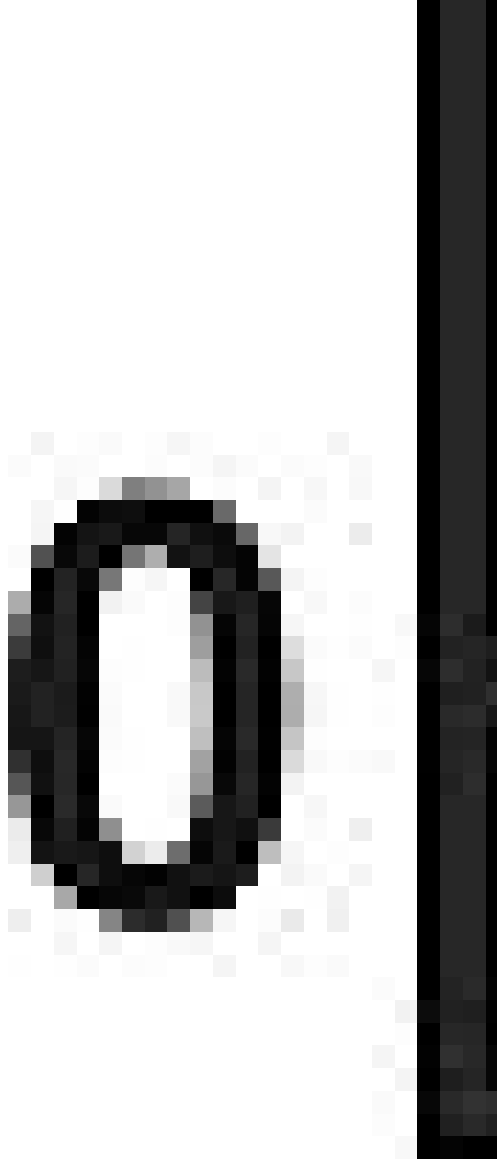}} \protect\protect\protect\protect\caption{$e_{2}\left(n\right)$ \eqref{eq:e2n} for $\boldsymbol{\Sigma}_{AR}$
\eqref{eq:AR} of AR(1) process with $p=50$ and $r=0.5$. }

\par\end{centering}

\begin{centering}
\label{icml-historical-2-1-2} 
\par\end{centering}

\vskip -0.2in 
\end{figure}

The related reconstruction error $e_{2}\left(n\right)$ \eqref{eq:e2n}
is depicted in Fig. 2. The reconstruction error $e_{2}\left(n\right)$
\eqref{eq:e2n} does not converge to zero due to its relative simplification
involving the use of $\tilde{\mathbf{F}}_{n}$ \eqref{eq:F-1} instead
of $\mathbf{F}_{n}$ \eqref{eq:F}. However, the use of $\tilde{\mathbf{F}}_{n}$
\eqref{eq:F-1} renders $\tilde{\boldsymbol{\Sigma}}_{2}^{-1}\left(\lambda_{n}\right)$
\eqref{eq:siginvapp-1} much more robust to outliers in comparison
to the first estimator $\tilde{\boldsymbol{\Sigma}}_{1}^{-1}\left(\lambda_{n}\right)$
\eqref{eq:siginvapp}. In that sense, a relatively small and fixed
reconstruction error can be assumed in order to avoid unexpected outliers.

\section{Conclusions}

A key challenge in many large-scale sequential machine learning methods
stems from the need to obtain the covariance matrix of the data, which
is unknown in practice and should be estimated. In order to avoid
additional complexity during the modeling process, it is commonly
assumed that the covariance matrix is known in advanced or, alternatively,
that simplified estimators are employed. In Section 3, we derived
a sequential update rule for the shrinkage estimator that offers a
compromise between the sample covariance matrix and a well-conditioned
matrix. The optimal shrinkage coefficient, in the sense of mean-squared
error, is analytically obtained, which is a notable advantage since
a key factor in realizing large-scale architectures is the resource
complexity involved. In Section 4, sequential update rules that approximate
the inverse shrinkage estimator are derived. The experimental results
in Section 5 clearly demonstrates that the reconstruction errors of
the approximated inverses are relatively small. The sequential update
rules that approximate the inverse of the shrinkage estimator provide
a general result that can be utilized in a wide range of sequential
machine learning applications. Therefore, the approach paves the way
for improved large-scale machine learning methods that involve sequential
updates in high-dimensional data spaces. 

\bibliographystyle{IEEEbib}
\bibliography{bib_ICML}

\end{document}